\documentclass[traditabstract]{aa} 
\newcommand{\Brak}[1]{{\left({#1}\right)}}
\newcommand{\D}{{\mathrm d}}
\newcommand{\SBrak}[1]{{\left[{#1}\right]}}
\newcommand{\DxDy}[2]{{\frac{\D{#1}}{\D{#2}}}}

\newcommand{\dxdycz}[3]{{\Brak{\frac{\partial{#1}}{\partial{#2}}}_{\!\!{#3}}}}
\newcommand{\Nad}{{\nabla_{\!\mathrm{ad}}}}
\newcommand{\Nmu}{{\nabla_{\!\mathrm{\mu}}}}
\newcommand{\N}{{\nabla}}

\newcommand{\vES}{{v_{\mathrm{ES}}}}
\newcommand{\vmu}{{v_{\mathrm{\mu}}}}
\newcommand{\vESO}{{v_{\mathrm{e}}}}
\newcommand{\HP}{{H_{\mathrm{P}}}}
\newcommand{\tauKHx}{{\tau^*_{\mathrm{KH}}}}
\newcommand{\Max}[1]{{\max\CBrak{#1}}}
\newcommand{\CBrak}[1]{{\left\{{#1}\right\}}}

\newcommand{\epsnu}{{\varepsilon_{\mathrm{\nu}}}}
\usepackage{graphicx}
\usepackage{txfonts}
\usepackage{natbib}
\bibpunct{(}{)}{;}{a}{}{,} 
\begin{document}
   \title{Numerical Tests of Rotational Mixing in Massive Stars with the new Population Synthesis Code BONNFIRES}

   \author{H.H.B. Lau
          \inst{1},
          R.G.Izzard\inst{1}
   \and F.R.N. Schneider\inst{1}}

   \institute{ Argelander-Institut f\"ur Astronomie, Universit\"at Bonn, Auf dem H\"ugel 71, D-53121 Bonn, Germany,  
              \email{hblau@astro.uni-bonn.de}
             }
   \date{Received 08 May 2014; accepted 19 Aug 2014}

  \abstract{We use our new population synthesis code BONNFIRES to test how surface abundances predicted by rotating stellar models depend on the numerical treatment of rotational mixing, such as spatial resolution, temporal resolution and computation of mean molecular weight gradients. In stellar evolution codes the process of transporting chemical species and angular momentum is usually approximated as a diffusion process. We find that even with identical numerical prescriptions for calculating the rotational mixing coefficients in the diffusion equation, different timesteps lead to a deviation of the coefficients  and hence surface abundances. We find the surface abundances vary by 10-100\% between the model sequences with short timestep of 0.001\,Myr to model sequences with long timesteps of 0.1-1\,Myr. Model sequences with stronger surface nitrogen enrichment also have longer main-sequence lifetimes because more hydrogen is mixed to the burning cores. The deviations in main-sequence lifetimes can be as large as 20\%. Mathematically speaking, no numerical scheme can give a perfect solution unless infinitesimally small timesteps are used, which is computationally not practical. However, we find that the surface abundances eventually converge within 10\% between modelling sequences with sufficiently small timesteps below 0.1\,Myr. The efficiency of rotational mixing depends on the implemented numerical scheme and critically on the computation of the mean molecular weight gradient. A smoothing function for the mean molecular weight gradient results in stronger rotational mixing. When comparing observations with detailed theoretical models made by stellar evolutionary codes or population synthesis codes such as BONNFIRES, deviations of surface abundances because of numerical treatments have to be considered carefully. Calibrations of rotational mixing parameters therefore depend on the chosen discretization schemes. If the discretization scheme or the computational recipe for calculating the mean molecular weight gradient is altered, re-calibration of mixing parameters may be required to fit observations. If we are to properly understand the fundamental physics of rotation in stars, it is crucial that we minimize the uncertainty introduced into stellar evolution models when numerically approximating rotational mixing processes.}

\authorrunning{H.H.B.Lau, R.G.Izzard, F.R.N. Schneider}
\titlerunning{Numerical Tests of Rotational Mixing}
\maketitle
%

\section{Introduction}

One pressing problem in modern stellar astronomy
is whether rotation induces mixing in stars and hence alters their surface abundances. Rotationally induced mixing processes can prolong stellar lifetimes and have been used to explain the observed surface enrichment of the hydrogen burning products, such as nitrogen, in some massive main-sequence stars. How rotation affects stellar evolution and nucleosynthesis has been recently reviewed by \cite{Langer2012} and \cite{Maeder2012}. However, the interaction between rotation and binary mass transfer or with magnetic fields remains uncertain, so quantitative tests of rotational mixing in single stars are vital particularly if we want to understand the role of rotation in more complex stellar systems.

In the Large Magellanic Cloud sample of the VLT-FLAMES survey \citep{Evans2011}, observed B stars with both rapid rotation and nitrogen enrichment support the underlying theory of rotationally induced mixing \citep{Hunter2008}. However, the interpretation of the Hunter diagram, which shows the observed surface nitrogen abundances as a function of the projected rotational velocity, is complicated by the significant fractions of nitrogen-enhanced slow-rotators and nitrogen-normal rapid-rotators. These stars cannot easily be explained by current theories of single rotating massive stars \citep{Brott2011b}, although it is argued that the stellar parameters derived for some of the observed stars may need revision \citep{Maeder2014}. \citet{deMink2013} suggest a significant fraction, or perhaps even close to all of the rapid rotators, are the products of close binary evolution. Based on a sample of 68 massive stars in our Galaxy, \citet{Aerts2014} also deduce that neither projected rotational velocity nor the rotational frequency has predictive power over the measured nitrogen abundance and hence any mixing cannot be because of rotational mixing alone. Accurate and consistent predictions for rotational mixing are thus crucial to understand the role of rotation in stellar evolution.

There are various numerical prescriptions of rotational mixing  \citep[e.g.][]{Heger,Maeder2000} and they all contain two or more uncertain parameters controlling the mixing efficiency. Usually these uncertain parameters are calibrated against observed surface enrichments. For example, \citet{Brott2011a} calibrate the efficiency of rotational mixing, $f_{\mathrm{c}}$, and the parameter $f_\mu$, which parametrizes the stabilizing effect of the mean molecular weight gradient, against the observed B stars in the FLAMES survey. These parameterizations are crucial to quantitatively model how surface abundances vary with rotational velocity. In this work we demonstrate that this parameterization is only valid for the chosen numerical scheme.

While other rotational instabilities such as shear instabilities \citep[e.g.][]{Zahn1974} can transport chemical species, we focus on the Eddington-Sweet circulation \citep[][]{Sweet1950,Endal1978} and the Goldreich-Schubert-Fricke (GSF) instability \citep[][]{Goldreich1967,Fricke1968} and how the two instabilities are stabilized by the presence of a mean molecular weight gradient. In a main-sequence massive star, the mean molecular weight gradient is particularly relevant to mixing across the inner convective core with higher mean molecular weight to the radiative envelope with lower mean molecular weight. In this work, we investigate how surface nitrogen enrichments by these two rotational instabilities depend on the mean molecular weight computation.

In stellar evolution codes the transport process of chemical species and angular momentum is usually approximated as a diffusion process. In some codes, advection is included for the transport of angular momentum \citep{Maeder2012}. Discretizations in mass and time are required to solve the continuous diffusion equation. Discretizing in mass corresponds to cutting the star into thousands of meshpoints while discretizing in time corresponds to dividing the evolutionary sequences into timesteps when solving the diffusion equation. The diffusion coefficient is then calculated based on physical properties at each meshpoint at the start of the timestep. However, because the abundance profile is always evolving within any single timestep, any calculated value of the rotational mixing coefficient based on the chemical profile and mean molecular weight gradient at a particular time, usually the start of the timestep, cannot be strictly correct for the whole timestep. As a result even with identical numerical prescriptions and mixing parameters, e.g. $f_{\mathrm{c}}$, model sequences with different timesteps deviate in rotational mixing coefficients and hence surface abundances. The deviations are bigger for large timesteps especially near the start or the end of rotational instabilities when the coefficient changes from zero to non-zero or vice versa. 

To make the situation worse, rotational mixing at the current timestep feeds back on the abundance profile at the next timestep. If the diffusion coefficient is overestimated in one timestep, too much material is mixed which leads to a smaller mean molecular gradient and hence more mixing in next timestep. The divergence of chemical profiles can run away over the course of the evolutionary model sequence because of this positive feedback. 

The numerical treatment and computation of the mean molecular weight gradient are also critical for rotational mixing. Particularly at the edge of the burning core in a moderately fast rotating star, rotationally induced mixing and mean molecular weight gradient induced stability can be similar. Therefore, if a different prescription for the computation of the mean molecular weight gradient is used, the rotational mixing coefficient changes. 

This poses a serious problem for the calibration of uncertain rotational mixing parameters. Although the parameters are well calibrated against observations for a given discretization scheme, drastically different results can be obtained if different timesteps or meshpoints are used in stellar models. Therefore, when comparing observations to theoretical models, one needs to be aware of the hidden uncertainty caused by the choice of the numerical scheme.   

In theory only model sequences with infinitesimally small timesteps converge to the identical results. However it is computationally not practical for any code to use an infinitesimally small timesteps. In practice the model sequences start to converge if sufficiently small timesteps are used. We estimate the minimal timesteps required to achieve an accurate result such that the deviations of the surface abundances because of insufficient temporal resolution are kept within 10\% of our model with the best resolution. We also show that the chemical compositions and the efficiency of rotational mixing depend critically on how mean molecular weight gradients are computed. If the computational recipe for the mean molecular weight gradient is altered, re-calibration of mixing parameters may be required to fit observations.

This numerical problem exists in other similar mixing processes where the calculation of diffusion coefficients depends on the mean molecular weight gradient. For example, thermohaline mixing is driven by the inversion of the mean molecular gradient. In low-mass red giants, the depletion of lithium by thermohaline mixing strongly depends on the timestep used in detailed stellar evolutionary codes (Lattanzio et al. 2014, in prep). 

In this paper, we use a new population synthesis code BONNFIRES to run grids of model sequences to study quantitatively how rotational mixing depends on adopted timesteps and shell masses. BONNFIRES is a new population synthesis code designed to model the internal mixing and nucleosynthesis of binary stars in a timely fashion. \citet[][]{Sana2012,Sana2013} show that almost all massive stars are born in a binary system, so significant fractions of stars observed are expected to have undergone binary interaction to reach their current states \citep{deMink2014}. One of the major aims of BONNFIRES is to investigate whether binary interactions can account for the observed nitrogen-rich slow-rotators and nitrogen-normal fast-rotators.

\section{BONNFIRES}

BONNFIRES is a population synthesis code designed to model the internal mixing and nucleosynthesis of binary stars in a timely fashion with detailed internal composition profiles, mixing processes and nuclear reaction networks.  Internal structure variables such as temperature, density, pressure, and opacity are interpolated from input models produced by the Binary Evolution Code BEC \citep{Heger,Petrovic2005bec,Yoon2006}. Each input model has 1000 meshpoints regular spaced in mass, so for a $20M_{\odot}$ model, our mass resolution is limited to $0.02M_{\odot}$. Our input models are non-rotating and have metallicity $Z=0.02$. The abundance pattern is chosen according to \citet{Grevesse1993}. We find nearly identical evolution of surface compositions even when we use models with initial equatorial rotational velocity $300\,\rm{km\,s^{-1}}$ as input models, so we conclude that the structural feedback by rotation for moderately fast rotators up to an initial equatorial rotational velocity of  $300\,\rm{km\,s^{-1}}$ is minimal with regards to the problems we address. Other sets of input models may be required for faster rotations, e.g. those displaying chemically homogeneous evolutions although this is outside the scope of this work. 

We treat mass loss by stellar winds by implementing the prescription of \citet{Vink2001}. Rotational mass loss enhancement is included according to
\citet{Yoon}. Angular momentum is removed assuming the mass lost carries the same specific angular momentum as the surface shell. These mass-loss treatments are the same as the underlying input BEC models. Treatments of mass loss are not important for this work because only around $0.5M_{\odot}$ is lost during the main-sequence of the $20M_{\odot}$ star considered.

Our stellar models are discretized into “mesh points of mass $\delta m$. The angular velocity, chemical abundances and moment of inertia of each shell are defined in the middle while other physical quantities such as radius and mass are defined at the inner edge of the mass shell. The different definitions are not important as long as the shell mass is sufficiently small. We do not include any convective overshooting and adopt the Ledoux criterion for convection, so a stratification is stable against convection \citep{Kippenhahn1990} if
\begin{equation}
  \Nad-\N+\frac{\varphi}{\delta}\Nmu \geq 0\,,
\end{equation}
where
\begin{equation}
\Nad=\dxdycz{\,\ln T}{\,\ln P}{S,X}
\;,\quad
\Nmu=\DxDy{\,\ln \mu}{\,\ln P}
\;,\quad
\N=\DxDy{\,\ln T}{\,\ln P}\,
\; \textrm{and}
\end{equation}
\begin{equation}
\delta=-\dxdycz{\ln\rho}{\ln T}{\mu,P}
\;,\qquad
\varphi=\dxdycz{\ln\rho}{\ln\mu}{P,T}
\;,
\end{equation}
where $T$ is the temperature, $P$ is the pressure, $\mu$ is the mean molecular weight, $\rho$ is the density and the index ``ad'' means adiabatic, i.e. at constant entropy and composition. An equation of state is included in BONNFIRES to calculate the above gradients (Glebbeek, private communication). In the models shown in this work, the convective mixing coefficients are read from input models. Convective zones are located for regions where the convective mixing timescales are smaller than the timesteps, and those regions are assumed to be instantly mixed and hence have a homogeneous composition.

We follow \citet{Heger} in our treatment of rotational mixing and the calculation of mixing coefficients based on the abundance profile and structure variables. Mixing coefficients based on shear instabilities are also computed according to the prescription by \citet{Heger}, although we find that the contribution from shear instabilities is insignificant compared to Eddington-Sweet circulation in our model sequences.

 We follow \citet{Brott2011a} for the values of the rotational mixing parameters. Chemical mixing is treated as a diffusive process and
implemented by solving the following diffusion equation

\begin{equation}
  \dxdycz{X_n}{t}{m}=\dxdycz{}{m}{t}\; \SBrak{\,(4\pi r^2 \rho)^2  D\,
     \dxdycz{X_n}{m}{t}\,}
\,,
\end{equation}
where $D$ is the sum of diffusion coefficients of all
individual mixing processes, including convection and all rotational mixing processes;  $X_n$ is the mass fraction of species
$n$; $m$ is the Lagrangian mass co-ordinate of the shell with radius $r$; and $t$ is the time. 

The abundances are calculated independently by a nuclear reaction network while the diffusion equation is solved by a diffusion solver. This is called operator splitting in the literature. The pp, CNO and NeNaMgAl chains are solved for using a Kaps-Rentrop scheme \citep{Press2002} and NACRE reaction rates \citep{NACRE}.  We accurately reproduce the nucleosynthesis in the detailed code models. We solve the diffusion equation with the Euler method. The compositions are iterated with an implicit diffusion solver until the profile satisfies the diffusion equation used for that timestep. This method is sometimes referred to as the implicit finite difference method \citep{MM2004}.

The diffusion coefficient is calculated based on the composition profile at the start of a timestep. However, while the solver is implicit, this mixing scheme is not totally implicit because only one coefficient, calculated at the start of the timestep and assumed to be constant, is used for the whole timestep. This is not precise because, in reality, the coefficient varies with the composition profile within each timestep. For the rest of our discussion, we label this original scheme as explicit to match the explicit nature of the diffusion coefficient. In our explicit scheme, the diffusion coefficient D for a timestep $\delta t$ always follows the equation

\begin{equation}
D(t)=D(t_{0})=D_{0}\,\,\,\,\, t_{0}\leq t<t_{0}+\delta t\,,
\label{eq:explicit}
\end{equation}
where $D_{0}$ is the diffusion coefficient at the start of the timestep $\delta t$.

We also employ an alternative implicit scheme that takes into account the variation of the mixing coefficient throughout the timestep. We assume the coefficient varies linearly within each timestep and hence the diffusion coefficient is the average of the coefficient at the start and at the end. In the first iteration, the new diffusion coefficient $D_{1}=0.5\times(D_{0}+D_{\mathrm{f}0})$, where the coefficient $D_{\mathrm{f}0}$ is calculated at the end of the timestep after solving the diffusion equation with initial diffusion coefficient $D_{0}$. For the next iteration, $D_{1}$ is used as the diffusion coefficient which results in a new chemical profile at the end of the timestep and hence a slightly different coefficient $\mathrm{D_{\mathrm{f}1}}$. The diffusion coefficient at the $n$th iteration is

\begin{equation}
D_{n}=0.5\times(D_{0}+D_{\mathrm{f} (n-1)})\,,
\label{eq:implicit}
\end{equation}
where  $D_{\mathrm{f} (n-1)}$ is the coefficient calculated at the end of the timestep in the previous $(n-1)$th iteration.

We iterate until the new coefficient $D_{n}$ agrees with the coefficient of the previous iteration $D_{n-1}$ within a threshold 10\% in all meshpoints.  We use this scheme to minimize deviation between model sequences with different timesteps.

 \begin{figure}

   \includegraphics[width=\columnwidth,angle=0]{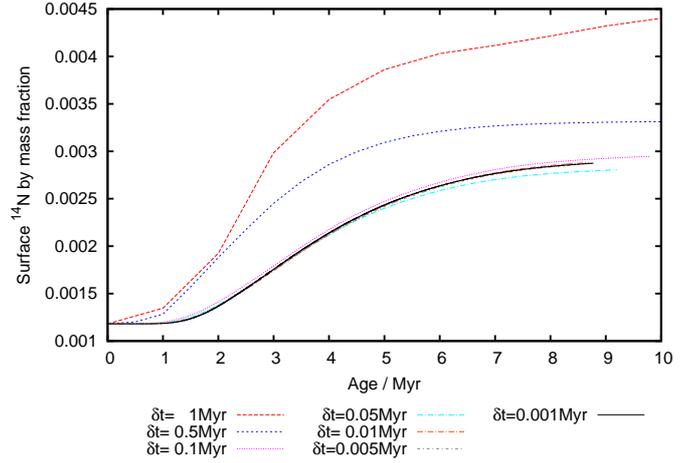}
   \caption{Surface $^{14}\mathrm{N}$ abundance as a function of stellar age of our model $20M_{\odot}$ star  with initial equatorial rotational velocity of  $300\,\rm{km\,s}^{-1}$. Model sequences are run with fixed shell mass $\delta m=0.05M_{\odot}$  with different timesteps $\delta t$.  These model sequences are run with BONNFIRES using the implicit scheme for calculating diffusion coefficients, as described in Section 2. We note that deviations of surface abundances may fluctuate non-monotonically, for example, the model sequence with $\delta t=0.05\,$Myr has slightly lower surface  $^{14}\mathrm{N}$ abundances than sequences with smaller timesteps.   }
 \label{fig:fixedtime}

   \end{figure}

   \begin{figure}
   
   \includegraphics[width=\columnwidth,angle=0]{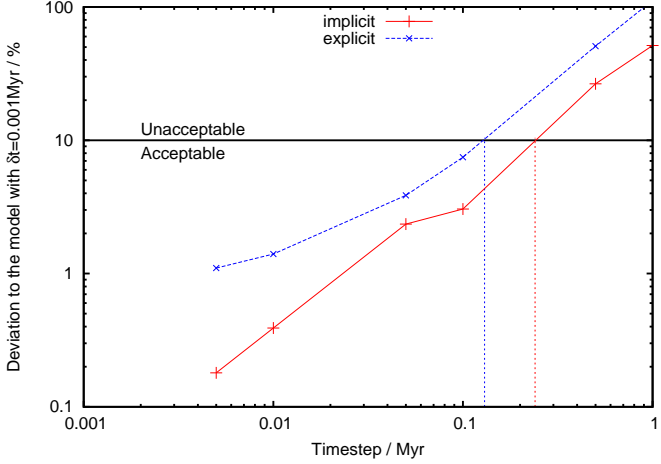}
   \caption{Percentage deviations between our implicit (red solid) and explicit (blue dotted) diffusive mixing schemes with different timesteps. The deviations are taken for nitrogen abundances at the time of the evolution when the differences between that evolutionary sequence with the sequence with fixed timestep $\delta t$=0.001\,Myr is the largest. The model sequences are for a $20M_{\odot}$ star with initial equatorial rotational velocity of  $300\,\rm{km\,s}^{-1}$ with fixed shell mass $\delta m=0.05M_{\odot}$. 
    }
\label{fig:error}
   \end{figure}

\subsection{Models}

We follow the evolution of a single star until the end of the main-sequence which we define when the central hydrogen mass fraction $X_{\mathrm{H}}=10^{-5}$. The only mixing mechanisms in our models are convection and rotational mixing. The
initial mass of our model sequences is $20M_{\odot}$ with an initial equatorial rotational velocity of  $300\,\rm{km\,s}^{-1}$. All model sequences are run with fixed timesteps  $\delta t$ and fixed shell masses  $\delta m$ throughout their evolution. We vary the timesteps and shell masses
to investigate how surface abundances change based on different mass and temporal resolution. The timesteps  $\delta t$ range from 0.001\,Myr to 1\,Myr and the shell masses  $\delta m$ range from  $0.02M_{\odot}$ to  $0.5M_{\odot}$. 

Detailed stellar model sequences have a variable timestep throughout the evolution and shell masses throughout the star. For example, our input stellar models are calculated with suitable mass resolution so that the atmosphere is solved for correctly. Because we do not need to solve the stellar structure equations in BONNFIRES and seek to model interior mixing and nucleosynthesis rather than the stellar atmosphere, we do not have as many meshpoints in the atmosphere to increase the speed of population synthesis calculations. For our resolution test, it is much simpler to compare surface abundances between model sequences with different shell masses and timesteps if the timesteps and shell masses are constant in each sequence.

\section{Results}

We find that the surface abundances of elements such as nitrogen depend sensitively on the
adopted timestep in the model sequences. Moreover, different mass resolution
and a different computation method of the mean molecular weight gradient can lead to significant differences in the predicted nitrogen abundances and main-sequence lifetimes.

\subsection{Timestep dependency}

  \begin{figure}
   
   \includegraphics[width=\columnwidth,angle=0]{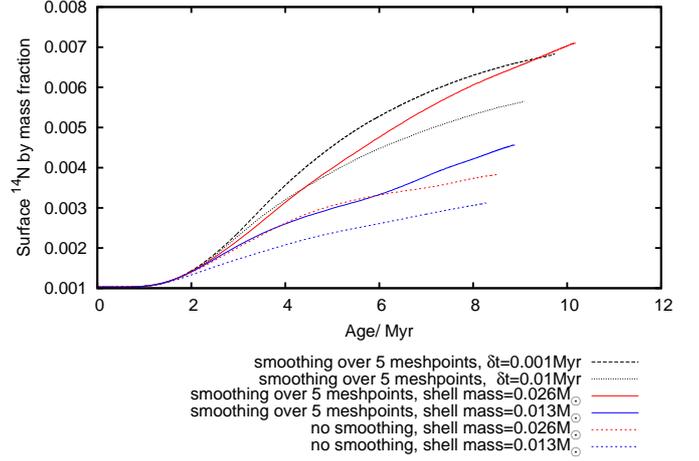}
   \caption{Surface $^{14}\mathrm{N}$ abundance as a function of stellar age of our model $20M_{\odot}$ star with initial equatorial rotational velocity of  $300\,\rm{km\,s}^{-1}$ produced by the stellar evolution code BEC. The surface nitrogen abundance deviates between the two otherwise identical model sequences, with different fixed timesteps $\delta t$=0.001\,Myr (black dashed) and $\delta t$=0.01\,Myr (black dotted). The main-sequence lifetime is also different by about 10\% between both model sequences. Numerical smoothing of the gradient (solid lines) enhances mixing of nitrogen to the surface. Models with higher resolution (blue lines) have a larger mean molecular weight gradient at the boundary of burning region which prohibits mixing. For models without fixed timesteps, the timesteps $\delta t$ are adjusted by BEC code automatically and are of the order of 0.001\,Myr.}
\label{fig:becnitrogen}
   \end{figure}

Figure~\ref{fig:fixedtime} shows how the evolution of surface nitrogen mass fraction in our model sequences varies as a function of timestep. With our implicit scheme, in which the coefficient is taken as the average at the start and at the end of the timestep, nitrogen converges for model sequences with small timesteps but it deviates significantly in model sequences with timesteps larger than 0.1\,Myr. Furthermore, the main-sequence lifetimes are longer for model sequences with larger nitrogen enhancements because rotational mixing simultaneously mixes hydrogen to the core as well as nitrogen to the surface. Without the implicit averaging scheme, the nitrogen abundance evolution varies even more with timesteps $0.1$\,Myr or longer (Fig.~\ref{fig:error}). In the extreme case of $\delta t$=1\,Myr the nitrogen abundance is doubled. Only model sequences with timesteps of 0.05\,Myr or shorter agree within a few per cent.

We define the deviation of each model sequence as the maximum deviation of surface nitrogen abundances between each model sequence to the model sequence with the shortest timestep (0.001\,Myr) throughout the evolution. In general, the deviations are smaller by 50\% or more with the implicit averaging scheme (Fig.~\ref{fig:error}). However, our implicit averaging scheme (Eq.~\ref{eq:implicit}) is computationally more expensive for each evolutionary model because the number of iterations is typically around ten. The total computational time for the whole evolutionary sequence with our implicit scheme is comparable with using the explicit scheme (Eq.~\ref{eq:explicit}) with a shorter timestep to achieve similar accuracy. For example, using a timestep of 0.05\,Myr with our explicit scheme is slightly computationally faster than using a timestep of 0.1\,Myr with the implicit scheme, while the percentage errors are comparable in both cases. 
Therefore, for the purpose of population synthesis, the implicit scheme is not necessary unless we need an accuracy within 1\%. Without the implicit looping, the error is still more than 1\% even for the model with timestep 0.005\,Myr. Nevertheless, an implicit scheme averaging the diffusion coefficients within a single timestep is worth considering in a detailed evolutionary code to achieve greater accuracy when the bottleneck in computation time is, rather, solving the stellar structure equations. For the purpose of population synthesis of stars around $20M_{\odot}$, we recommend a timestep shorter than 0.1\,Myr for the study of rotational mixing such that the estimated errors are within 10\%. The timestep required for a star more massive than $20M_{\odot}$ is likely to be even shorter.

We use the stellar evolution code BEC to run two $20M_{\odot}$ model sequences with initial equatorial rotational velocity of $300\,\rm{km\,s}^{-1}$ with different timesteps to illustrate that the timestep dependency is not unique to our population synthesis code BONNFIRES. The timesteps are fixed to be 0.01\,Myr and 0.001\,Myr respectively. We find around 10\% differences in the nitrogen abundance between the model sequences (Fig.~\ref{fig:becnitrogen}). The model sequence with stronger nitrogen enrichment also has a 10\% longer main sequence lifetime because more hydrogen is mixed to the core burning region by rotation. Because of a different numerical scheme and parameterization of rotational mixing, the BEC code results cannot be directly compared with BONNFIRES results. Nevertheless, the two model sequences show that detailed evolutionary codes suffer the same numerical problems for rotational mixing as our new population synthesis code, BONNFIRES. 

 \begin{figure}

   \includegraphics[width=\columnwidth,angle=0]{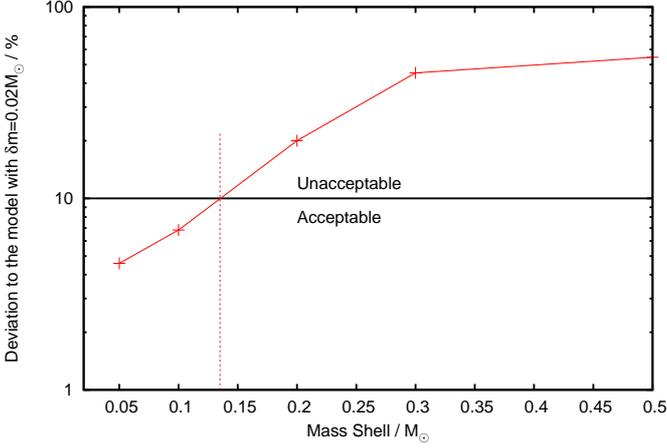}
  
   \caption{Percentage deviations for BONNFIRES model sequences with fixed shell masses $\delta m$. The deviations are taken for nitrogen abundances at the time of the evolution when the difference between that evolutionary sequence against sequence with fixed $\delta m$= $0.02M_{\odot}$ is the largest. The model sequences are for a $20M_{\odot}$ star with initial equatorial rotational velocity of  $300\,\rm{km\,s}^{-1}$ with fixed timestep $\delta t$=0.1\,Myr.}
\label{fig:fixeddm}

   \end{figure}

   \begin{figure}

   \includegraphics[angle=270,width=1.07\columnwidth,]{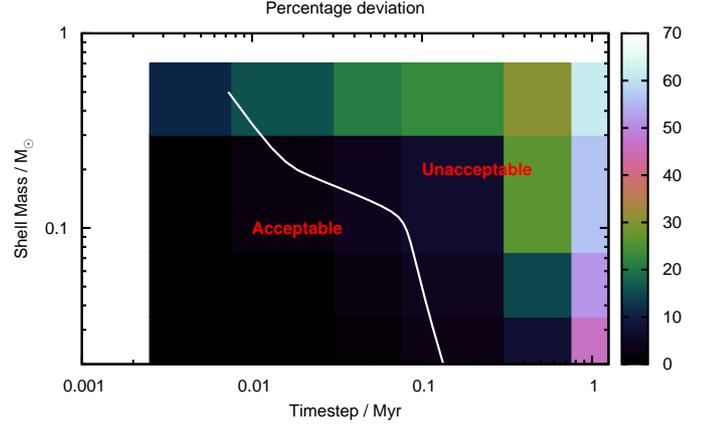}
   \caption{Percentage deviations between BONNFIRES evolutionary sequences with varying fixed timesteps $\delta t$ and shell masses $\delta m$ from the sequence with the smallest timestep $\delta t$=0.001\,Myr and the smallest shell mass $\delta m = 0.02M_{\odot}$. The deviations are taken for $^{14}\mathrm{N}$ abundances when the percentage deviation is the largest during the whole evolutionary sequence. The line indicates the minimal resolution required such that deviations are below 10\%.   }
\label{fig:2Derror}

   \end{figure}

\subsection{Dependence on mass resolution and computation of mean molecular weight}

With BONNFIRES, accuracy problems because of large shell masses are not as dramatic as those because of large timesteps, but the model with a fixed $0.5M_{\odot}$ shell mass still introduces a significant deviation of around 20\%. With our implicit scheme, a choice of fixed $0.1M_{\odot}$ shell mass with fixed timestep of 0.1\,Myr agrees within 10\% to model sequences of fixed $0.02M_{\odot}$ shell mass
with fixed timestep of 0.1\,Myr (Fig.~\ref{fig:fixeddm}).

Fig.~\ref{fig:2Derror} shows the percentage deviations of all BONNFIRES evolutionary sequences with various timesteps $\delta t$ and shell masses $\delta m$ compared to the sequence with the shortest timestep $\delta t$=0.001\,Myr and the smallest shell mass $\delta m = 0.02M_{\odot}$. Any model sequences with longer timesteps than $\delta t$=0.3\,Myr have large deviations irrespective of the shell mass. Similarly, the percentage errors are much larger for model sequences with shell mass larger than $0.3M_{\odot}$. The solid white line indicates the boundary of evolutionary sequences with less than 10\% deviations.

The above analysis only considers numerical resolution in the models. In addition, we find that rotational mixing depends on how the mean molecular weight gradient is calculated. For the Eddington-Sweet circulation and GSF instability, the mean molecular weight gradient can be written as a stabilizing circulation velocity (Eq. 36 of \citealt{Heger}), 
\begin{equation}
\vmu=\frac{\HP}{\tauKHx}
      \frac{\varphi f_{\mu}\Nmu}{\delta\,\Brak{\nabla-\Nad}}\,,
\label{eq:vmu}
\end{equation}
where $\HP$ is the pressure scale-height; $\tauKHx$ is the local Kelvin-Helmholtz time-scale; and  $\varphi$, $\Nmu$, $\delta$, $\nabla$, $\Nad$ are defined as before. The parameter $f_{\mu}$ describes the sensitivity of the rotationally induced mixing to $\mu$-gradient. We take $f_{\mu}=0.1$ as in \citet{Brott2011a}.
The pressure scale-height  $\HP$ is defined by
\begin{equation}
\HP=-\DxDy{r}{\ln P}
  =
     \frac{P}{\rho g}\, ,
\end{equation}
where $g$ is local gravity and the local Kelvin-Helmholtz timescale $\tauKHx$ is defined as
\begin{equation}
\tauKHx=\frac{G m^2}{r \Brak{l - m \epsnu}}\, ,
\end{equation}
where $l$ is luminosity and $\epsnu$ is defined as the energy generation rate due to neutrino losses and therefore is negative. Neutrino losses are insignificant in our main-sequence stellar models. The effective circulation velocity $\vES$ is reduced by stabilizing currents due to $\mu$-gradients  (\citealt{Endal1978} and Eq. 38 of \citealt{Heger}),

\begin{equation}
\vES=\Max{\bigl| {\vESO}\bigr| - \bigl| {\vmu}\bigr| ,0}\,,
\end{equation}
where $\vESO$ is the original circulation velocity (Eq. 35 of \citealt{Heger}).

In a moderately rapidly rotating star with initial equatorial rotational velocity around $300\,\rm{km\,s^{-1}}$, $\bigl|{\vESO}\bigr|$ and $\bigl|{v_\mu}\bigr|$ can be of similar magnitude, particularly at the edge of the burning core. The bottleneck for rotational mixing across the whole star is the location where  $\bigl|{\vmu}\bigr|$ is biggest which is at the edge of the convective core because of the composition discontinuity between convective core and radiative envelope. If there is numerical smoothing of the mean molecular weight or its gradient, which is inevitable in a discretized stellar model, the mean molecular weight gradient is averaged out and hence reduced in magnitude. Therefore, the maximum stabilizing term is reduced and rotational mixing is more efficient compared to models with no numerical averaging. In the detailed evolution code BEC, there is an option to take a weighted average of the mean molecular weight gradient across several meshpoints. In the $20M_{\odot}$ star with initial equatorial rotational velocity of $300\,\rm{km\,s^{-1}}$, we compare model sequences in which the gradients are averaged over five points, which is used in previous models by \citet{Brott2011b}, with model sequences in which no smoothing is used and find their surface nitrogen abundances are inconsistent (Fig.~\ref{fig:becnitrogen}). The mean molecular weight gradient across the convective core boundary also depends on the resolution because of the steep abundances variation between the burning and non-burning regions. As the resolution increases, the maximum mean molecular weight gradient increases and hence the efficiency of the rotational mixing drops. \citet{Mestel1953} suggests that near the convective core mixing can be impeded by a large negative molecular weight gradient and the Eddington-Sweet circulation can be divided into two separate zones by the discontinuity in composition. If this is the case, the stellar surface is not enriched by any materials from the burning core. However because the two separate zones can rotate at a different angular velocity, the shear instability \citep{Endal1978} may eventually mix materials across the two zones if sufficient differential rotation builds up.

  \begin{figure}

   \includegraphics[angle=270,width=\columnwidth]{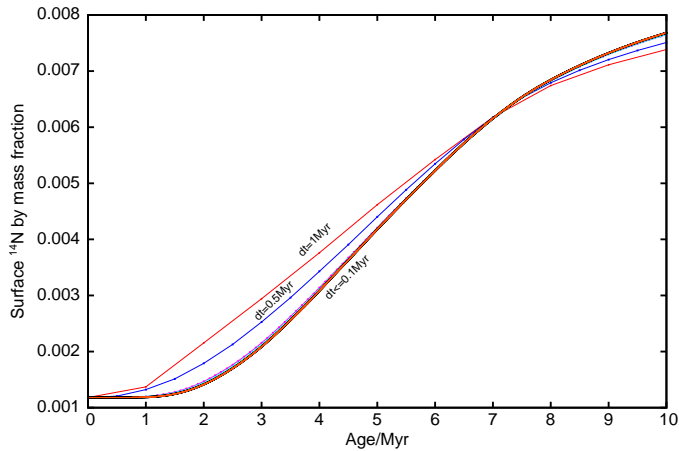}
   \caption{Surface $^{14}\mathrm{N}$ abundance as a function of stellar age of our model $20M_{\odot}$ star with initial equatorial rotational velocity of  $300\,\rm{km\,s}^{-1}$. Rotational mixing coefficients are calculated without any inhibitance of the mean molecular weight gradient.
    Models sequences have a $0.05M_{\odot}$ fixed shell mass with various timesteps in Myr by BONNFIRES. The surface nitrogen enhancements are similar for model sequences with all timesteps from 0.001\,Myr to 0.1\,Myr. }
 \label{fig:lednodiv}
   \end{figure}

BONNFIRES has artificial numerical smoothing because it linearly interpolates between input grids of models with 1000 regular meshpoints. While we find that a mesh spacing of $0.1M_{\odot}$ limits the deviation because of mass resolution within 10\% in BONNFIRES, we caution that this may not necessarily work in other detailed evolutionary codes because the mean molecular weight gradient can be steeper at the edge of the convective core than in our interpolated models.

\section{Discussion}

Results from BONNFIRES show that a time-implicit diffusion scheme (Eq.~\ref{eq:implicit}) improves the accuracy of surface abundances by a factor of two or more. With BONNFIRES, the maximum timestep for less than 10\% deviation in surface abundances is 0.1\,Myr, although this limit may not apply for other detailed evolutionary codes because of their differing numerical schemes. Moreover, stars more massive than $20M_{\odot}$ may require a shorter timestep. These types of numerical issues lead to inaccuracies in the calibration of mixing parameters and predicted surface compositions. Even if the same code with the same mixing parameters is used, different timesteps can lead to greater than $10$\% deviations if the timesteps are not sufficiently small. We have shown that thick shells also lead to an inaccuracy in the calculation of rotational mixing. However, we do not need thin shells throughout the whole star, for example, more massive shells can be used deep inside the core. Only at the edge of the convective core, where the mean molecular weight changes rapidly with mass, are more meshpoints required to resolve the variation in mean molecular weight to improve the accuracy of rotational mixing.

\citet{MM2004} test three numerical methods for solving the diffusion equation for the transport
of chemical species and angular momentum. Their models with large timesteps agree with each other in the implicit finite difference method, but the diffusion coefficients are kept constant as a function of time in their numerical experiments. We also adopt the implicit finite difference method, so the large abundance deviations for long timesteps in our models cannot be attributed to any numerical artifacts of solving the diffusion equation. Their results are consistent with our suggestion that abundances deviate at large timesteps because diffusion coefficients vary with time.

To test whether the mean molecular weight gradient is the major reason for the divergence of chemical profiles,
we run a set of model sequences with the BONNFIRES without the mean molecular weight gradient as inhabiting force by setting $f_{\mu}=0$, which is equivalent to setting $\vmu=0$ in Eq.~\ref{eq:vmu}. We find that in these model sequences, the nitrogen enhancements are increased relative to models with finite $\vmu$ and more importantly the nitrogen abundances in our model sequence with large timestep (0.5\,Myr) agrees to within 5\% of model sequences with smaller timesteps (Fig.~\ref{fig:lednodiv}). This shows that the mean molecular weight gradient is the major culprit for the numerical problems described.

As shown in the previous section, our results depend on the applied prescription for the computation of the mean molecular weight gradient. Averaging leads to increased rotational mixing because the mean molecular weight barrier is smoothed out. In the Eddington-Sweet circulation and GSF instability,
a sharp mean molecular weight barrier efficiently stabilizes against the
circulation. If that is the case, the abundance profile at the edge of the
convective core significantly influence the efficiency of rotational mixing, and hence the subsequent surface abundance profile.
In fact, rotational mixing between the core and the envelope may depend on the smoothing length even more than
the prescription for the rotational mixing itself. A realistic mixing scale length based on the stellar structure instead of an arbitrary number of meshpoints will improve consistency. However, it is not clear what this physical scale length should be and 3D
hydrodynamical simulations with rotation may be required to investigate how rotational mixing operates at the interface between the convective core
and the radiative envelope, and to determine the scale length.

As discussed in \citet{Mestel1953}, at the edge of the convective zone the sharp mean molecular weight gradient may inhabit any circulation and hence there is no flow across the composition barrier. However, other mixing mechanisms such as convective overshooting across the convective boundary may physically smoothen the mean molecular weight gradient. For example, a convective overshooting \citep[e.g.][]{Hollowell1988} prescription similar to \citet{Herwig2000}, in which the convective diffusion coefficient decays exponentially from the convective boundary, smooths the chemical profile. Adopting a similar smoothing scheme should reduce the dependency of the rotational mixing efficiency on the numerical computation of the mean molecular weight gradient at the cost of an extra free parameter to determine the magnitude of the overshooting. 

If a star rotates rapidly enough, the stabilizing effect of the mean molecular weight gradient will usually be overcome. This efficiently mixes between the core and the surface regardless of the numerical scheme. However, the onset of homogeneous evolution clearly depends on the computation of the mean molecular weight gradient and whether overshooting is adopted, as well as mass and temporal resolution. Therefore, the minimum equatorial rotational velocity required for homogeneous evolution probably also depends on the numerical scheme adopted.

Our study of the numerical treatment of rotational mixing on surface abundances is relevant to detailed stellar evolution codes, e.g. BEC \citep{Yoon2006} and MESA \citep{MESA}, that adopt a similar prescription to \citet{Heger}. Other rotational mixing prescriptions such as those used by \citet{Maeder2000} have a different prescription for calculating the rotational mixing coefficient, in particular the mean molecular weight gradient is smoothed by horizontal turbulence. \citet{Maeder98} have shown that the impact of horizontal turbulence on the efficiency of the transport of chemicals by the meridional currents can be important in reducing the stabilizing effect of the mean molecular weight gradients. Therefore, results from BONNFIRES (and, by extension, other similar codes) may not be valid for other rotational mixing prescriptions that include the impact of horizontal turbulence on transport by meridional circulation \citep{Chaboyer92}. Similarly to convective overshooting, the inclusion of horizontal turbulence reduces the stabilization of mean molecular weight gradient although there is debate on the efficiency of horizontal turbulence. Horizontal turbulence leads to more substantial surface helium enhancements which distinguish if any processes can mix materials across the mean molecular weight gradient barrier, although currently there are not enough observed helium abundances to draw any conclusion \citep{Langer2012}.  

Altogether, these numerical issues pose a serious problem for the calibration of uncertain mixing parameters. Although such parameters have already been calibrated against observations \citep[e.g.][]{Brott2011a}, different parameters may be obtained from re-calibration if alternative numerical schemes such as different timesteps or computation methods of the mean molecular weight gradients are employed. Therefore, when comparing observations with theoretical models, one needs to be aware of these hidden, but quantitatively crucial, uncertainties. \citet{Aerts2014} suggest the simultaneous action of internal gravity waves and differential rotation to be the best explanation for the measured surface nitrogen abundances in massive stars. Large-scale low-frequency internal gravity waves excited by the convective core, as in the studies by e.g. \citet{Neiner2012}, \citet{Rogers2012} and \citet{Mathis2013}, may contribute considerably to the surface nitrogen abundance if these waves are able to transport chemical species and angular momentum \citep{Rogers2013}. Minimizing the uncertainties of rotational mixing because of numerical treatments is crucial to quantitatively understand the role of rotation with other possible mixing mechanisms.

\section{Conclusion}
We have shown that chemical mixing by rotation has a significant dependence on timesteps, shell masses and numerical schemes used in the calculation of stellar evolution model sequences. This numerical issue leads to uncertainties in the predicted surface chemical enhancements by rotational mixing and also predicted minimal rotational velocities for chemically homogeneous evolution. Using our population synthesis code BONNFIRES, surface $^{14}\mathrm{N}$ abundances agree within 10\% in model sequences with timesteps shorter than 0.1\,Myr and shell masses smaller than 0.1$M_{\odot}$ for a $20M_{\odot}$ star with initial equatorial rotational velocity $300\,\rm{km\,s^{-1}}$. In order to achieve the same precision, the timestep required for a star more massive than $20M_{\odot}$ is likely to be shorter. In detailed evolutionary codes the timesteps and shell masses may need to be smaller because of their different numerical schemes. We find that the evolution of the mean molecular weight profile is the major reason for the numerical inaccuracy. The abundance profile, and hence the mean molecular weight gradient, is always evolving within any single timestep so any calculated value of the rotational mixing coefficient based on the chemical profile at a particular time cannot be strictly correct for the whole timestep. Therefore, we suggest an implicit scheme for calculating the diffusion coefficient such that it is the average between the coefficient at the start and at the end of each timestep. This averaging scheme can halve the deviation in surface abundances compared to only using the coefficient calculated at the start for the whole timestep. In other mixing processes that depend on the mean molecular weight gradient, such as thermohaline mixing and semi-convection, we suggest our scheme (Eq.~\ref{eq:implicit}) may also give a more precise result than simple time-explicit diffusive mixing.

The efficiency of rotational mixing also strongly depends on the computation of the mean molecular weight gradient. At the edge of the convective zone, the presence of a sharp mean molecular weight gradient stabilizes rotational mixing, so any schemes that smoothen the mean molecular weight or its gradient lead to stronger rotational mixing and hence stronger surface enrichment of hydrogen burning products and longer main-sequence lifetime. Because the efficiency of rotational mixing strongly depends on how strong the mean molecular weight gradient inhibits the flow across the composition barrier, hydrodynamical simulations with rotation may be required to investigate how rotational mixing operates at the interface between the convective core and the radiative envelope.

\begin{acknowledgements}
RGI and HBL thank the Alexander von Humboldt Foundation for support for this work. We thank Evert Glebbeek for the subroutine to calculate equation of state and Norbert Langer for useful discussion. We also thank the anonymous referee for suggestions to improve the manuscript.
\end{acknowledgements}

\bibliographystyle{aa} 
\bibliography{HBL.bib} 

\end{document}